\documentclass[12pt]{article}

\usepackage{cite}
\usepackage{epsfig}

\input paperdef

\newcommand{\epc}[2]{{\em Eur. Phys. J.}            {\bf C#1}, #2 }

\newcommand{\etal}{{\em et al.}}

\newcommand{\as}{\alpha_s}

\begin{document}
\thispagestyle{empty}

\def\thefootnote{\fnsymbol{footnote}}

\begin{flushright}
CERN--TH/2000--109\\
DESY 00--050\\
KA--TP--5--2000\\
UPR--882--T\\
hep-ph/0005024 \\
\end{flushright}

\vspace{0.5cm}

\begin{center}

{\large\sc {\bf Physics Impact of GigaZ}}

\vspace{0.8cm}

{\sc 
J.~Erler$^{1}$%
, S.~Heinemeyer$^{2}$%
, W.~Hollik$^{3}$%
, G.~Weiglein$^{4}$%
~and P.M.~Zerwas$^{2}$%
}

\vspace*{0.8cm}

{\sl
$^1$ Department of Physics and Astronomy, University of Pennsylvania, \\
Philadelphia PA 19104--6396, USA

\vspace*{0.4cm}

$^2$ DESY, Deutsches Elektronen-Synchrotron, D--22603 Hamburg, Germany

\vspace*{0.4cm}

$^3$ Institut f\"ur Theoretische Physik, Universit\"at Karlsruhe, \\
D--76128 Karlsruhe, Germany

\vspace*{0.4cm}

$^4$ CERN, TH Division, CH-1211 Geneva 23, Switzerland

\vspace*{0.4cm}
}

\end{center}

\vspace*{0.5cm}

\begin{abstract}
By running the prospective high-energy \epem collider TESLA in the
GigaZ mode on 
the $Z$~resonance, experiments can be performed on the basis of more than 
$10^9~Z$~events. They will allow the measurement of the effective
electroweak mixing angle to an accuracy of 
$\de\sweff \approx \pm 1 \times 10^{-5}$. 
Likewise the $W$~boson mass is expected to be measurable with an error of 
$\de\MW \approx \pm 6$~MeV near the $W^+W^-$ threshold. In this note, we study 
the accuracy with which the Higgs boson mass can be determined from 
loop corrections to these observables in the Standard Model. The comparison 
with a directly observed Higgs boson may be used to constrain new physics 
scales affecting the virtual loops. We also study constraints on the heavy 
Higgs particles predicted in the Minimal Supersymmetric Standard Model, 
which are very difficult to observe directly for large masses. Similarly, 
it is possible to constrain the mass of the heavy scalar top particle.
\end{abstract}

\def\thefootnote{\arabic{footnote}}
\setcounter{page}{0}
\setcounter{footnote}{0}

\newpage


\section{Introduction}

The prospective high-energy \epem\ linear collider TESLA is being
designed to operate on top of the $Z$~boson resonance,  
\BE
e^+\,e^- \to Z \, ,
\EE
by adding a bypass to the main beam line~\cite{gigaZ}.
Given the high luminosity, $\cL = 7 \times 10^{33} \rm{cm}^{-2} \rm{s}^{-1}$, 
and the cross section, $\si_Z \approx 30 \rm{~nb}$~\cite{Zxsection} 
(including radiative corrections), about $2 \times 10^9~Z$~events can be 
generated in an operational year of $10^7 \rm{s}$. We will therefore 
refer to this 
option as the GigaZ mode of the machine. Moreover, by increasing the collider 
energy to the $W$-pair threshold,
\BE
e^+\,e^- \to W^+ W^- \, ,
\EE
about $10^6$ $W$~bosons can be generated at the optimal energy point for 
measuring the $W$~boson mass, $\MW$, near threshold 
and about $3 \times 10^6$ $W$~bosons at the energy of maximal cross 
section~\cite{Wxsection}. The large increase in the number of $Z$~events by 
two orders of magnitude as compared to LEP~1  and the 
increasing precision in the measurements of $W$~boson properties, open new 
opportunities~\cite{gigazsitges} for high precision physics in the electroweak
sector~\cite{ewtheory1,ewtheory2}. 

By adopting the Blondel scheme~\cite{blondel} for running \epem\
colliders with polarized beams, the left-right asymmetry,
$A_{LR} \equiv 2 (1 - 4 \sweff)/(1 + (1 - 4 \sweff)^2)$,
can be measured with very high precision, 
$\de A_{LR} \approx \pm 10^{-4}$~\cite{moenig}, when both electrons and 
positrons are polarized longitudinally. This accuracy can be achieved
since the total cross section, the left-right asymmetry  and  the polarization
factor, $\cP = (P_+ + P_-)/(1 + P_+ P_-)$, can be measured 
by individually flipping the electron and positron helicities, generating 
all $2 \times 2$ spin combinations in $\si_{ij} (i,j = L,R)$; 
only the difference between the moduli $|P_+|$ and $|P_-|$ 
before and after flipping the polarizations of both the positron and
electron beams
need to be monitored by laser Compton scattering. From $A_{LR}$ the
mixing angle  
in the effective leptonic vector coupling of the on-shell $Z$~boson,
$\sweff$, can be determined to an accuracy 
\BE
\de\sweff \approx \pm\,1 \times 10^{-5},
\EE
while the $W$~boson mass is expected to be measurable to a precision of
\BE
\label{delmw}
\de\MW \approx \pm\,6 \mev,
\EE
by scanning the $W^+ W^-$ threshold~\cite{wwthreshold}.

Besides the improvements in $\sweff$ and $\MW$, GigaZ has the potential 
to determine the total $Z$~width within $\de\Gamma_Z = \pm 1$~MeV; the ratio 
of hadronic to leptonic partial $Z$~widths with a relative uncertainty of 
$\de R_l/R_l = \pm 0.05 \%$; the ratio of the $b\bar{b}$ to the hadronic 
partial widths with a precision of $\de R_b = \pm 1.4 \times 10^{-4}$; and 
to improve the $b$ quark asymmetry parameter $A_b$ to a precision of
$\pm 1 \times 10^{-3}$~\cite{moenig,gigazsitges}. 
These additional measurements offer complementary information on 
the Higgs boson mass, $\MH$, but also on 
the strong coupling constant, $\als$, 
which enters the radiative corrections in many places. 
This is desirable in 
its own right, and in the present context it is important to control $\as$ 
effects from higher order loop contributions to avoid confusion with Higgs 
effects. Indirectly, a well known $\as$ would also help to control $\mt$ 
effects, since $\mt$ from a threshold scan at a linear collider will be 
strongly correlated with $\as$. 
We find that via a precise measurement of $R_l$, GigaZ would provide
a clean determination of 
$\as$ with small error 
\BE
\de\as \approx \pm\, 0.001 , 
\label{delalphas}
\EE
allowing to reduce the error of the top-quark mass from the threshold
scan. 
The anticipated precisions for the most relevant electroweak observables 
at the Tevatron (Run IIA and IIB), the LHC, a future linear collider, LC,
and GigaZ are summarized in \refta{tab:precallcoll}. 

\begin{table}[ht!]
\renewcommand{\arraystretch}{1.5}
\BC
\begin{tabular}{|c||c||c|c|c||c|c|}
\cline{2-7} \multicolumn{1}{c||}{}
           & now & Tev.\, Run IIA & Run IIB & LHC & LC  & GigaZ \\ \hline
                                                                      \hline
$\de\sweff(\times 10^5)$& 17 & ~~~~~~~50 \hfill~\cite{Baur96} &
                               ~~~13 \hfill~\cite{Baur96} &
                               ~~21 \hfill~\cite{Baur96,EWrep} &
                               ~~(6) \hfill~\cite{Baur96} &
                               ~1.3 \hfill~\cite{moenig}\\ \hline
$\de\MW$ [MeV] & 42          & ~~~~~~~30 \hfill~\cite{Signore98} &
                               ~~~15 \hfill~\cite{Keller97} &
                               ~~15 \hfill~\cite{Keller97,EWrep} &
                               ~~15 \hfill~\cite{Haber97} & 
                               ~~6 \hfill~\cite{wwthreshold}\\ \hline
$\de\mt$ [GeV] & 5.1         & ~~~~~~~4.0 \hfill~\cite{Baur96} &
                               ~~2.0 \hfill~\cite{Baur96} &
                               ~2.0 \hfill~\cite{Baur96,Toprep} &
                               ~0.2 \hfill~\cite{Frey97} &
                               0.13 \hfill~\cite{Frey97}\\ \hline
$\de\MH$ [MeV] & ---         & --- &
                               2000 \hfill~\cite{Gunion97} &
                               100 \hfill~\cite{Gunion97} &
                               ~~50 \hfill~\cite{Gunion97} &
                               ~50 \hfill~\cite{Gunion97}\\ \hline
\end{tabular}
\renewcommand{\arraystretch}{1}
\caption[]{\it\footnotesize
Expected precision at various colliders for $\sweff$, $\MW$, $\mt$  and 
the (lightest) Higgs boson mass, $\MH$, at the reference value $\MH = 110$~GeV.
Run IIA refers to 2~fb$^{-1}$ integrated luminosity per experiment collected 
at the Tevatron with the Main Injector, while Run IIB assumes the accumulation 
of 30~fb$^{-1}$ by each experiment. LC corresponds to a linear
collider without the GigaZ  
mode. (The entry in parentheses assumes a fixed target polarized M\o ller 
scattering experiment using the $e^-$ beam.) 
The present uncertainty on $\MW$ will be improved by the end of the
LEP2 program.
$\de\mt$ from the Tevatron and 
the LHC is the error for the top pole mass, while at the top threshold in 
\epem\ collisions the \msbar\ top-quark mass can be determined. The smaller
value of $\de\mt$ at GigaZ compared to the LC is due to the prospective
reduced uncertainty of $\alpha_s$, which affects the relation between
the mass parameter directly extracted at the top threshold and the \msbar\
top-quark mass.
}
\label{tab:precallcoll}
\EC
\end{table}

In this note, we study the potential impact of such measurements on
the parameters of the Standard Model (SM) and its minimal
supersymmetric extension (MSSM)~\cite{mssm}.  
Higgs boson masses and SUSY particle masses affect the high precision
observables  through loop corrections. These loop corrections are evaluated 
in this note at the presently available level of theoretical accuracy, 
still leaving many refinements to be worked out in the coming 
years~\cite{loopverein}. Even though a complete set of calculations is 
lacking at the present time, the essential features of the 
GigaZ physics potential can nevertheless be studied in first exploratory
steps. In \citere{gigazsitges} it has been demonstrated that very
stringent consistency tests of the SM and the MSSM will become feasible
with the GigaZ precision, and the prospects for $b$~physics 
at GigaZ have been discussed. The latter topic has been studied in more 
detail in \citere{moenig}.
In the present note, we will focus in a systematic way
on the Higgs sectors of the SM and 
the MSSM, and also on the scalar top sector of the MSSM.


\section{Higgs Sector of the SM}

In the canonical form of the SM, the precision observables measured at 
the $Z$~peak are affected by two high mass scales in the model:
the top quark mass, $\mt$, and the Higgs boson mass, $\MH$. 
They enter as virtual states in loop corrections to various 
relations between electroweak observables. For example, 
the radiative corrections entering the relation between 
$\MW$ and $\MZ$, and between $\MZ$ and $\sweff$, have a strong quadratic 
dependence on $\mt$ and a logarithmic dependence on $\MH$. We mainly
focus on the two electroweak observables that are expected to be measurable 
with the highest accuracy at GigaZ, $\MW$ and $\sweff$. Our analysis
is based on the results for the electroweak precision observables
including higher order electroweak~\cite{sm2lmt4,sm2lnl} and
QCD~\cite{sm2lqcd,sm3lqcd} corrections. The current theoretical
uncertainties~\cite{PCP} are dominated by the parametric uncertainties 
from the errors in the input parameters $\mt$ (see \refta{tab:precallcoll}) 
and $\Delta\alpha$. The latter denotes the QED-induced shift
in the fine structure constant, $\al \to \al(\MZ)$, originating from 
charged-lepton and light-quark photon vacuum polarization diagrams. 
The hadronic contribution to $\De\al$ currently introduces an uncertainty of 
$\de\De\al = \pm 2 \times 10^{-4}$~\cite{delalphatheorydriven}. Forthcoming 
low-energy \epem annihilation experiments may reduce this uncertainty to 
about $\pm 5 \times 10^{-5}$~\cite{delalphajegerlehner}. 
Combining this value with future (indistinguishable) errors from
unknown higher order corrections, we assign the total uncertainty
of $\de\De\al = \pm 7 \times 10^{-5}$ to $\De\al$,
which is used throughout the paper unless otherwise stated.
For the future theoretical uncertainties from unknown higher-order
corrections (including the uncertainties from $\de\De\al$)
we assume,
\BE
\delta\MW(\mbox{theory}) = \pm 3 \mev, \quad
\delta\sweff(\mbox{theory}) = \pm 3 \times 10^{-5} \quad
\mbox{(future)} .
\label{eq:futureunc}
\EE
Given the high precision of
GigaZ, also the experimental error in $\MZ$, 
$\de\MZ = \pm 2.1 \mev$~\cite{delMZ}, results in non-negligible
uncertainties of $\de\MW = \pm 2.5 \mev$ and 
$\de\sweff = \pm 1.4 \times 10^{-5}$.
The experimental error in the top-quark mass, $\de\mt = \pm 130 \mev$,
induces further uncertainties of $\de\MW = \pm 0.8 \mev$ and 
$\de\sweff = \pm 0.4 \times 10^{-5}$. Thus, while currently the
experimental error in $\MZ$ can safely be neglected, for the GigaZ
precision it will actually induce an uncertainty in the prediction of
$\sweff$ that is larger than its experimental error.

\smallskip
\noindent
{\bf (a)} The relation 
between $\sweff$ and $\MZ$ can be written as
\BE
\label{deltars}
  \sweff \cweff = \frac{A^2}{\MZ^2 (1-\Delta r_Z)}, 
\EE
where $A = [(\pi\alpha)/(\sqrt{2} G_F)]^{1/2} = 37.2805(2)$~GeV is 
a combination of two precisely known low-energy coupling constants, the   
Fermi constant, $\gf$, and the electromagnetic fine structure constant, 
$\al$. The quantity $\Delta r_Z$ 
summarizes the loop corrections, which at the \onel\ level can be
decomposed as 
\BE
\Delta r_Z  =  \Delta \al - \De\rho^{\rm t} +\De r_Z^{\rm H}
               + \cdots .
\label{deltarz}
\EE
The leading top contribution to the $\rho$ parameter~\cite{rhoparameter}, 
quadratic in $\mt$, reads
\BE
\De \rho^{\rm t} = \frac{3 G_F m_t^2}{8\pi^2\sqrt{2}} \, .
\EE
The Higgs boson contribution is screened, being logarithmic for large 
Higgs boson masses 
\BE
\De r_Z^{\rm H}  =   \frac{G_F \MW^2}{8\pi^2\sqrt{2}} 
                       \frac{1+9\sw^2}{3\cw^2}
                   \log \frac{\MH^2}{\MW^2} +  \cdots .
\EE

\smallskip
\noindent
{\bf (b)} 
An independent analysis can be based on the precise measurement
of $\MW$ near threshold. The $\MW$--$\MZ$ interdependence is given by,
\BEA
  \frac{\MW^2}{\MZ^2} 
  \left(1-\frac{\MW^2}{\MZ^2} \right) & = &
  \frac{A^2}{\MZ^2 (1-\Delta r)},
\label{eq:b}
\EEA
where the quantum correction $\Delta r$ has the \onel\ decomposition,
\BEA
\Delta r & = & \Delta \al - \frac{\cw^2}{\sw^2} \De\rho^{\rm t}
              + \De r^{\rm H} 
             + \cdots    , \\ 
\De r^{\rm H} & = & \frac{G_F \MW^2}{8\pi^2\sqrt{2}} \frac{11}{3}  
                  \log \frac{\MH^2}{\MW^2} +  \cdots ,
\EEA
with $\De\al$ and $\De\rho^{\rm t}$ as introduced above.

\begin{figure}[ht!]
\vspace{1.5em}
\begin{center}
\mbox{
\psfig{figure=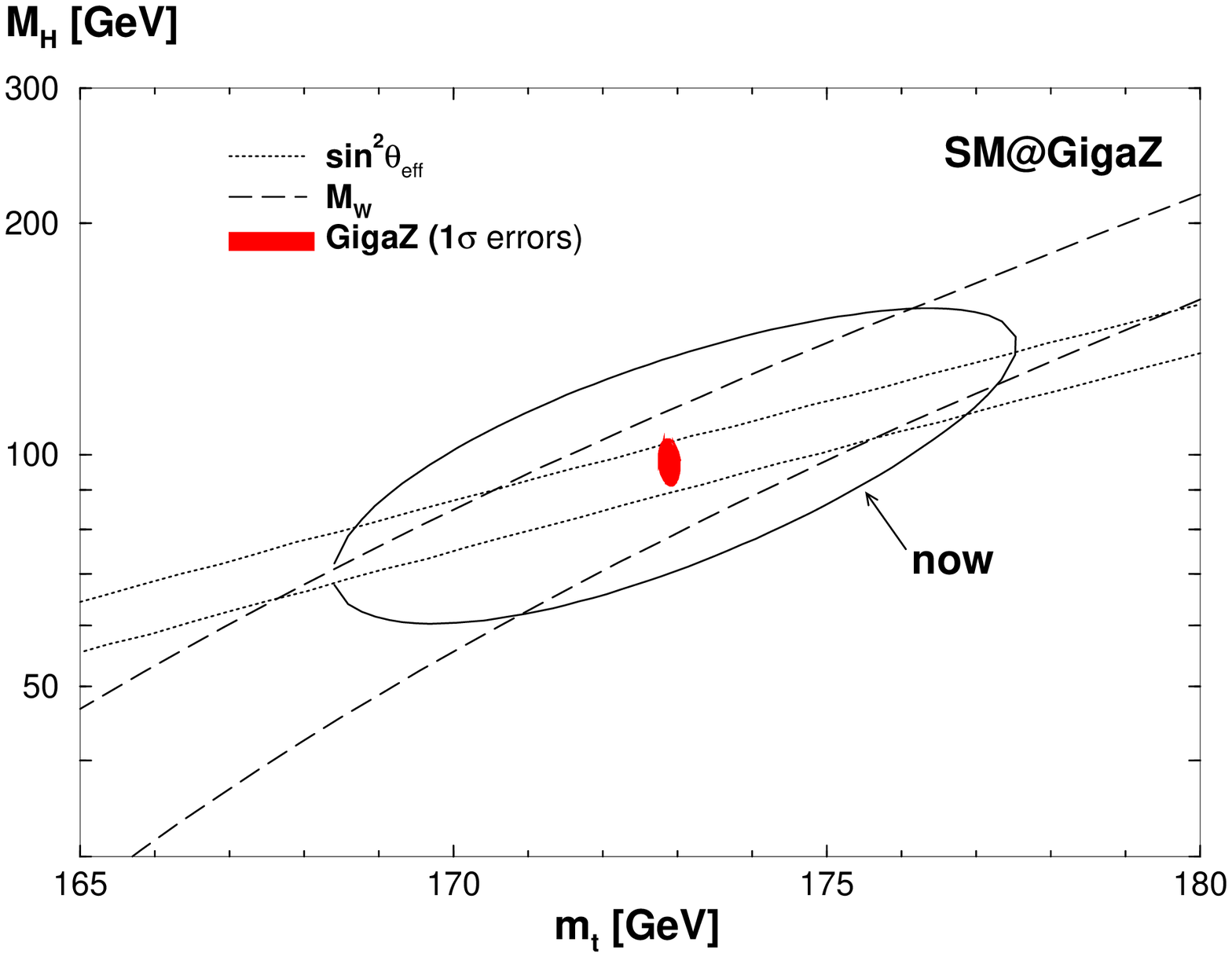,width=12cm,height=10.5cm}}
\end{center}
\caption[]{\it\footnotesize 
$1\sigma$ allowed regions in the $\mt$-$\MH$ plane taking into
account the anticipated GigaZ precisions
for $\sweff, \MW, \Ga_Z, R_l, R_q$ and $\mt$ (see text).
The presently allowed region (full curve labeled 'now') is shown for
comparison. 
}
\label{fig:mhmt}
\end{figure}

Owing to the different dependences of $\sweff$ and $\MW$ on $\mt$ and $\MH$,
the high precision measurements of these quantities at GigaZ (combined with 
the other supplementary electroweak observables) can determine the mass scales 
$\mt$ and $\MH$. The expected accuracy in the indirect determination of $\MH$ 
from the radiative corrections within the SM is displayed in \reffi{fig:mhmt}. 
To obtain these contours, the error projections in \refta{tab:precallcoll} 
are supplemented by central values equal to the current SM best fit values 
for the entire set of current high precision
observables~\cite{Erler99A}. For the theoretical uncertainties,
\refeq{eq:futureunc} is used, while the 
parametric uncertainties, such as from $\als$ and $\MZ$, are automatically
accounted for in the fits. 
The allowed bands in the $\mt$--$\MH$ plane for the GigaZ accuracy are
shown separately for $\sweff$ and $\MW$. By adding the information on the 
top-quark mass, with $\delta\mt \lsim 130$~MeV 
obtained from measurements of the $t\bar{t}$ production cross section near
threshold, an accurate determination of the Higgs boson mass becomes feasible
from both, $\MW$ and $\sweff$. If the two values are found to be consistent, 
they can be combined and compared to the Higgs boson mass measured in direct 
production through Higgs-strahlung~\cite{lep2higgs} 
(see the last row in \refta{tab:precallcoll}). 
In \reffi{fig:mhmt} this is shown by the shaded area labeled as
``GigaZ (1$\si$ errors)'', where the
measurements of other $Z$~boson properties as anticipated for GigaZ are
also included (the best fit value for $\mt$ is
assumed to coincide with the central $\mt$ value in \reffi{fig:mhmt}). 
For comparison, the area in the $\mt$--$\MH$ plane
corresponding to the current experimental accuracies, labeled as
``now (1$\si$ errors)'', is also shown.

The results can be summarized by calculating the accuracy with which $\MH$ 
can be determined indirectly. The expectations for $\de\MH/\MH$ in each step
until GigaZ are collected in \refta{tab:indirectmh}. Extending an
earlier analysis~\cite{gigazsitges}, where $\de\MH/\MH$ was separately
determined from $\MW$ and $\sweff$, we additionally employ here the
full set of precision observables for our analysis. Concerning the 
experimental errors, the values from \refta{tab:precallcoll} are taken. 
For the uncertainty in $\De\al$, we use
$\de\De\al = \pm 7 \times 10^{-5}$ (yielding \refeq{eq:futureunc} upon
combination with our estimate of unknown higher order corrections),
except for the first row displaying the present situation, where 
$\de\De\al = \pm 2 \times 10^{-4}$~\cite{delalphatheorydriven} is employed. 

It is apparent 
that GigaZ, reaching $\de\MH/\MH = \pm 7\%$, triples the precision in $\MH$ 
relative to the anticipated LHC status. On the other hand, a linear collider 
without the high luminosity option would provide only a modest improvement.

\begin{table}[htb!]
\renewcommand{\arraystretch}{1.5}
\BC
\begin{tabular}{|l||r|r||r|}

\hline
$\delta\MH/\MH$ {\it from:}            & $\MW$ & $\sweff$ & all  \\  \hline
                                                                     \hline
now                                    &229 \% & 62 \%    & 61 \% \\ \hline
                                                                     \hline
Tevatron Run IIA                       & 77 \% & 46 \%    & 41 \% \\ \hline
Tevatron Run IIB                       & 39 \% & 28 \%    & 26 \% \\ \hline
LHC                                    & 28 \% & 24 \%    & 21 \% \\ \hline
                                                                     \hline
LC                                     & 18 \% & 20 \%    & 15 \% \\ \hline
GigaZ                                  & 12 \% &  7 \%    &  7 \% \\ \hline
\end{tabular}
\renewcommand{\arraystretch}{1}
\caption[] {\it\footnotesize
{{\sl Cumulative\/} expected precisions of indirect Higgs mass
determinations, $\de\MH/\MH$, given the error projections in 
\refta{tab:precallcoll}. 
Theoretical uncertainties and their correlated effects on $\MW$ and $\sweff$ 
are taken into account (see text).
The last column shows the indirect Higgs mass determination from the
full set of precision observables.}   }
\label{tab:indirectmh}
\EC
\end{table}

\begin{figure}[ht!]
\vspace{1em}
\begin{center}
\mbox{
\psfig{figure=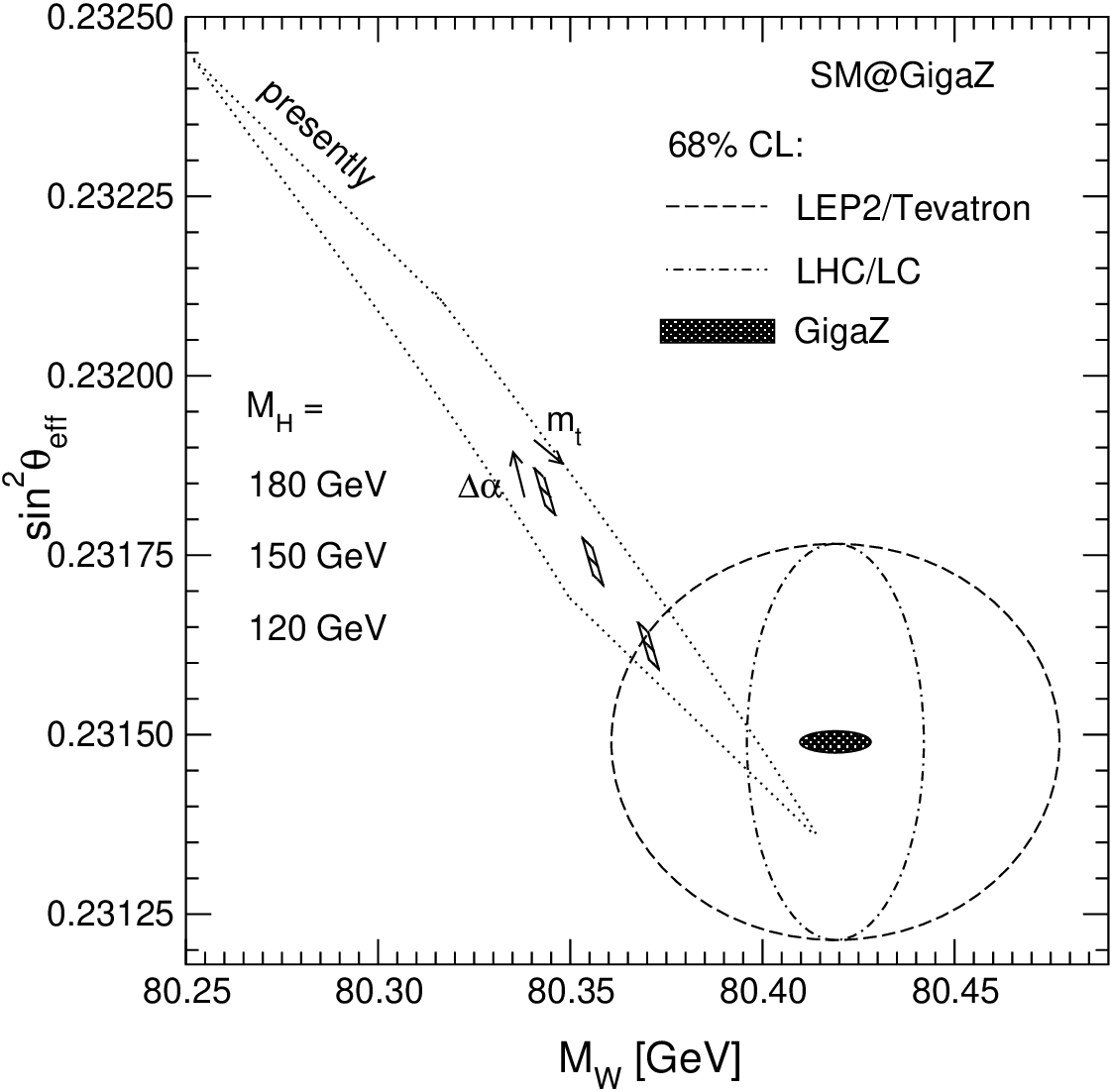,width=12cm,height=10cm}}
\end{center}
\caption[]{\it\footnotesize
The theoretical prediction for the relation between $\sweff$ and $\MW$
in the SM for Higgs boson masses in the intermediate range is compared to
the experimental accuracies at LEP~2/Tevatron (Run IIA), LHC/LC and GigaZ
(see \refta{tab:precallcoll}). For the
theoretical prediction an uncertainty of 
$\de\De\al = \pm 7 \times 10^{-5}$ and $\de\mt = \pm 200 \mev$ is taken 
into account.
}
\label{fig:SWMW}
\end{figure}

\smallskip
\noindent
{\bf (c)} 
A direct formal relation between $\MW$ and $\sweff$ can be established
by combining the two relations \refeqs{deltars} and (\ref{eq:b}) as
\BE
\label{deltar}
\MW^2 = \frac{A^2}{\sweff (1 - \De r_W)} .
\EE
The quantum correction $\De r_W$ is independent of $\De\rho^{\rm t}$
in leading order and has the \onel\ decomposition
\BEA
\De r_W  &=&  \De\al - \De r_W^{\rm H} + \cdots , \\
\De r_W^{\rm H}  &=&  \frac{G_F \MZ^2}{24\pi^2\sqrt{2}} 
                  \log \frac{\MH^2}{\MW^2} +  \cdots .
\label{deltarw}
\EEA
Relation~(\ref{deltar}) can be 
evaluated by inserting the measured value of the Higgs boson mass as 
predetermined at the LHC and the LC. This is visualized in \reffi{fig:SWMW},
where the present and future theoretical predictions for $\sweff$ and
$\MW$ (for different values of $\MH$) are compared with the experimental
accuracies at various colliders. Besides the independent predictions of
$\sweff$ and $\MW$ within the SM, the $\MW-\sweff$ contour plot in
\reffi{fig:SWMW} can be interpreted as an additional indirect
determination of $\MW$ from the measurement of $\sweff$.
Given the expected negligible error in $\MH$, this results in an
uncertainty of
\BE
\de\MW ({\rm indirect}) \approx \pm 2 \mev \pm 3 \mev .
\label{delmw2}
\EE
The first uncertainty reflects the experimental error in $\sweff$, while the 
second is the theoretical uncertainty discussed above (see
\refeq{eq:futureunc}). The combined uncertainty of this indirect
prediction is about the same as the one of the SM
prediction according to \refeq{eq:b}
and is close to the experimental error expected from the 
$W^+W^-$ threshold given in~\refeq{delmw}.

\smallskip
Consistency of all the theoretical relations with the experimental data would 
be the ultimate precision test of the SM based on quantum fluctuations. 
The comparison between theory and experiment can also be exploited to constrain
possible physics scales beyond the SM. These additional contributions can 
conveniently be described in terms of the S,T,U~\cite{Peskin90} or $\epsilon$
parameters~\cite{Altarelli90}. Adopting the notation of Ref.~\cite{Erler99A}, 
the errors with which they can be measured at GigaZ are given as follows:
\BE
\label{eq:STU}
\begin{array}{lr}
  \Delta S = \pm 0.05, & \hspace{50pt} \Delta \hat{\epsilon}_3 = \pm 0.0004, \\
  \Delta T = \pm 0.06, & \hspace{50pt} \Delta \hat{\epsilon}_1 = \pm 0.0005, \\
  \Delta U = \pm 0.04, & \hspace{50pt} \Delta \hat{\epsilon}_2 = \pm 0.0004.
\end{array}
\EE
The oblique parameters in \refeq{eq:STU} are strongly correlated.
On the other hand, many types of new physics predict $U = \hat\epsilon_2 = 0$ 
or very small (see ~\citere{Erler99A} and references therein). 
With the $U$ ($\hat\epsilon_2$) parameter known, the anticipated 
errors in $S$ and $T$ would decrease to about $\pm 0.02$, while the errors
in $\hat\epsilon_1$ and $\hat\epsilon_3$ would be smaller than $\pm 0.0002$.

In the context of a spontaneously broken gauge theory, the above mentioned
comparisons shed light on the basic theoretical components for generating 
the masses of the fundamental particles.  
On the other hand, an observed inconsistency would be a clear
indication for the existence of a new physics scale. 


\section{Supersymmetry}

The second step in this GigaZ analysis is based on the assumption that
supersymmetry would be discovered at LEP~2, the Tevatron, or the LHC, and 
further explored at an \epem\
linear collider. The high luminosity expected at TESLA can be
exploited to determine supersymmetric particle masses and mixing
angles with errors from \order{1\%} down to one per 
mille~\cite{susyattesla}, provided they reside in the kinematical reach of
the collider, which we assume to be about 1~TeV.
In this context we will address
two problems arising in the Higgs sector and the
scalar top sector within the MSSM.

For the SUSY contributions to $\MW$ and $\sweff$ we use
the  complete \onel\ results in the MSSM~\cite{deltarsusy} as well as
the leading higher order QCD corrections~\cite{mssm2lqcd}. The recent
electroweak two-loop results of the SM part in the MSSM~\cite{sm2lnl}
have not been taken into account, since no genuine MSSM counterpart is
available so far. As above, concerning the future theoretical 
uncertainties of $\MW$ and $\sweff$ we use \refeq{eq:futureunc}.

In contrast to the Higgs boson mass in the SM, the lightest $\cp$-even 
MSSM Higgs boson mass, $\Mh$, is not a free parameter but can be
calculated from the other SUSY parameters. In the present analysis, 
the currently most precise result based on Feynman-diagrammatic
methods~\cite{mhiggs2l} is used, relating $\Mh$ 
to the pseudoscalar Higgs boson mass, $\MA$. The
numerical evaluation has been performed with the Fortran code
\fh~\cite{feynhiggs}. In our analysis we assume a future uncertainty in 
the theoretical prediction of $\Mh$ of $\pm 0.5 \gev$.

\smallskip
\noindent
{\bf (a)} 
The relation between $\MW$ and $\sweff$ is affected by the parameters
of the supersymmetric sector, especially the
$\Stop$-sector. At the LHC~\cite{lhcstop} and 
especially at a prospective LC, 
the mass of the light $\Stop$, $\mste$, and the
$\Stop$-mixing angle, $\tst$, may be measurable very well, particularly
in the process $e^+\,e^- \to \Stope \bar{\Stope}$~\cite{lcstop}.
On the other hand,
background problems at the LHC and insufficient energy at the LC may
preclude the analysis of the heavy $\Stop$-particle,~$\Stopz$. 

\begin{figure}[ht!]
\vspace{1em}
\begin{center}
\epsfig{figure=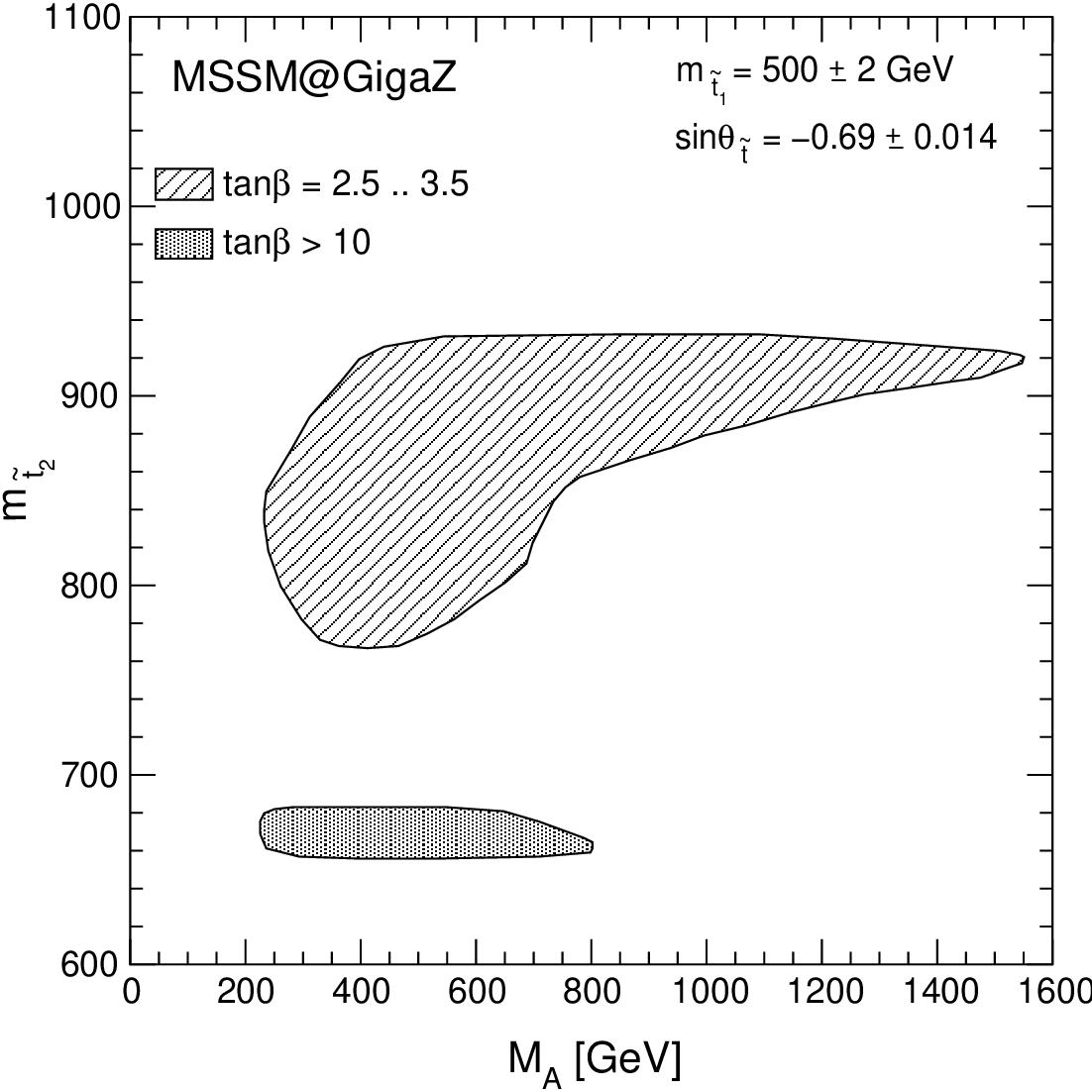,width=12cm,height=10cm}
\end{center}
\caption[]{\it\footnotesize
 The region in the $\MA-\mstz$ plane, allowed by
$1\,\si$ errors obtained from the GigaZ measurements of $\MW$ and $\sweff$: 
$\MW = 80.40 \gev$, 
$\sweff = 0.23140$, 
and from the LC measurement of $\Mh$:
$\Mh = 115 \gev$. 
The experimental errors for the SM parameters are given in
\refta{tab:precallcoll}. 
$\tb$ is assumed to be 
experimentally constrained by $2.5 < \tb < 3.5$ or $\tb > 10$. 
The other parameters including their uncertainties are given by 
$\mste = 500 \pm 2 \gev$,
$\sintt = -0.69 \pm 0.014$,
$\Ab = \At \pm 10\%$,
$\mu = -200 \pm 1 \gev$,
$M_2 = 400 \pm 2 \gev$ and
$\mgl = 500 \pm 10 \gev$.
For the uncertainties of the theoretical predictions we use
\refeq{eq:futureunc}.
}
\label{fig:MSt2MA}
\end{figure}

In \reffi{fig:MSt2MA} it is demonstrated how in this situation limits
on $\mstz$ can be derived from measurements of $\Mh$, $\MW$ and $\sweff$.
As experimental values we
assumed $\Mh = 115 \gev$, $\MW = 80.40 \gev$ and $\sweff = 0.23140$,
with the experimental errors 
given in the last column of \refta{tab:precallcoll}. 
We consider two cases for $\tb$, the ratio of the vacuum expectation values of 
the two Higgs doublets in the MSSM: the low $\tb$ region, 
where we assume a band, $2.5 < \tb < 3.5$, 
and the high $\tb$ region where we assume a lower bound, $\tb \geq 10$,
as can be expected from  
measurements in the gaugino sector (see e.g.\ \citere{tbmeasurement}).
As for the other parameters, the following values 
are assumed, with uncertainties as expected from LHC~\cite{lhctdr} and
TESLA~\cite{teslatdr}: 
$\mste = 500 \pm 2 \gev$,
$\sintt = -0.69 \pm 0.014$,
$\Ab = \At \pm 10\%$,
$\mu = -200 \pm 1 \gev$,
$M_2 = 400 \pm 2 \gev$  and
$\mgl = 500 \pm 10 \gev$.
($A_{b,t}$ are trilinear soft SUSY-breaking parameters,
$\mu$ is the Higgs mixing parameter, $M_2$ is one of the soft SUSY-breaking
parameter in the gaugino sector, and $\mgl$ denotes the gluino mass.)

For low $\tb$ the heavier $\Stop$-mass, $\mstz$, can be restricted to
$ 760 \gev \lsim \mstz \lsim 930 \gev$ from the $\Mh$, $\MW$ and $\sweff$ 
precision measurements. 
The mass $\MA$ varies between $200 \gev$ and $1600 \gev$. 
A reduction of this interval to $\MA \ge 500 \gev$ by its
non-observation at the LHC and the LC does not
improve the bounds on $\mstz$. 
If $\tb \ge 10$, the second theoretically
preferred range~\cite{su5so10}, the allowed region turns out to be
much smaller ($ 660 \gev \lsim \mstz \lsim 680 \gev$), and 
the mass $\MA$ is restricted to $\MA \lsim 800 \gev$.
In deriving the bounds on $\mstz$, both the constraints from $\Mh$ (see
\citere{tampprec}) and $\sweff$ play an important role. For the bounds
on $\MA$, the main effect comes from $\sweff$.
We have assumed a value for $\sweff$ slightly different from the 
corresponding value obtained in the SM limit.
For this value the (logarithmic) dependence on $\MA$ is 
still large 
enough so that in combination with the high precision in
$\sweff$ at GigaZ an {\em upper limit} on $\MA$ can be set.
For an error as obtained at an LC without the GigaZ mode (see
\refta{tab:precallcoll}) no bound on $\MA$ could be inferred.

\smallskip
\noindent
{\bf (b)} 
A similar problem of high interest occurs in the
sector of the MSSM Higgs particles. It is well known, that the heavy
Higgs bosons $A$, $H$  and $H^\pm$, are increasingly difficult to observe at
the LHC with rising mass~\cite{lhctdr}. At \epem\ linear
colliders heavy Higgs particles are produced primarily in pairs $(HA)$ and
$(H^+H^-)$ so that they cannot be analyzed 
for mass values beyond the beam
energy of $500 \gev$ in the first phase of such a machine. It has been
demonstrated though that the ratio of the decay branching 
ratios of the light Higgs boson $h$ is sensitive to $\MA$ up to 
values of
700~GeV to 1~TeV~\cite{battaglia}. Since any such analysis is difficult,
it is suggestive to search for complementary channels in which new
limits may be derived from other high precision measurements.

\begin{figure}[ht!]
\vspace{1em}
\begin{center}
\mbox{
\psfig{figure=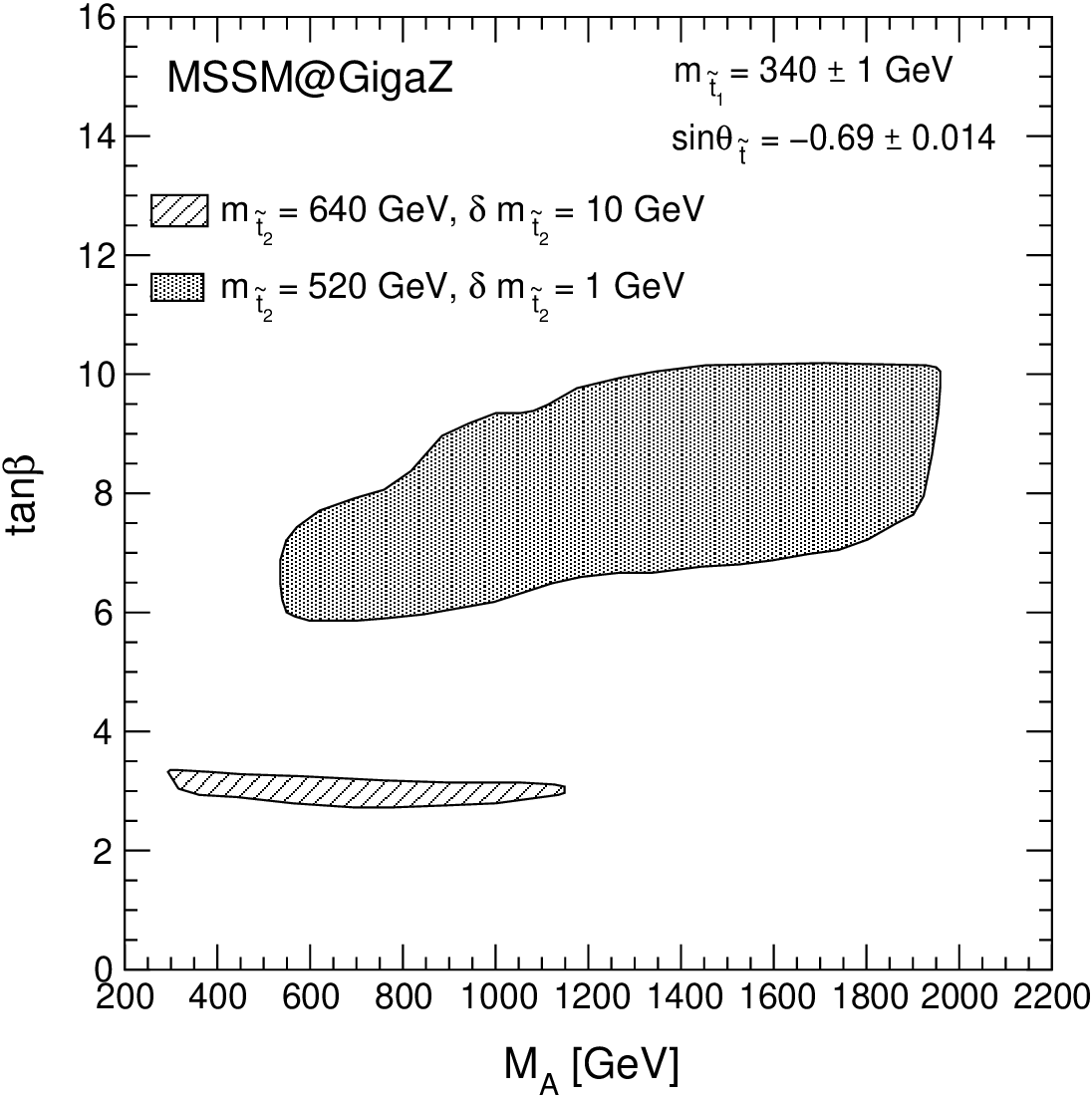,width=12cm,height=10cm}}
\end{center}
\caption[]{\it\footnotesize
 The region in the $\MA-\tb$ plane, allowed by
$1\,\si$ errors by the GigaZ measurements of $\MW$ and $\sweff$: 
$\MW = 80.40 \gev$, 
$\sweff = 0.23138$, 
and by the LC measurement of 
$\Mh$: $\Mh = 110 \gev$.
The experimental errors for the SM parameters are given in
\refta{tab:precallcoll}. 
The other parameters including their uncertainties are given by 
$\mste = 340 \pm 1 \gev$,
$\mstz = 640 \pm 10 \gev$ or 
$\mstz = 520 \pm 1 \gev$,
$\sintt = -0.69 \pm 0.014$,
$\Ab = -640 \pm 60 \gev$,
$\mu = 316 \pm 1 \gev$,
$M_2 = 152 \pm 2 \gev$ and
$\mgl = 496 \pm 10 \gev$.
For the uncertainties of the theoretical predictions we use
\refeq{eq:futureunc}.
}
\label{fig:TBMA}
\end{figure}

The result of such a study 
is presented in \reffi{fig:TBMA}, based on the expected errors for $\Mh$,
$\mste$, and $\tst$ from LC measurements, and assuming either a rough
measurement of the heavy $\Stop$-mass, $\mstz$, at the LHC, or
a precise determination of $\mstz$ at an LC.
\reffi{fig:TBMA} shows the
exclusion contours in the $\MA-\tb$ plane based on the following
scenario (inspired by the mSUGRA(1) reference scenario studied e.g.\ in
\citere{susyattesla}):
$\Mh = 110 \pm 0.05 \gev$ from LC measurements,
$\MW = 80.400 \pm 0.006 \gev$ and
$\sweff = 0.23138 \pm 1 \times\,10^{-5}$ from GigaZ measurements,
$\mste = 340 \pm 1 \gev$,
$\sintt = -0.69 \pm 0.014$ from the LC,
$\mstz = 640 \pm 10 \gev$ from the LHC or alternatively
$\mstz = 520 \pm 1 \gev$ from LC measurements; furthermore
$\Ab = -640 \pm 60 \gev$,
$\mu = 316 \pm 1 \gev$, 
$M_2 = 152 \pm 2 \gev$,
$\mgl = 496 \pm 10 \gev$ based on LHC or LC runs.

If the scenario with lower $\Stop$ masses is realized, this
would, due to the dependence of the Higgs boson mass on the MSSM
parameters, correspond to higher values for both $\MA$ and $\tb$.
Despite the precise measurement of $\mstz$ up to 
$1 \gev$ at the LC, in the example considered here
the restrictions placed on the Higgs sector would
be relatively weak. The constraints would be
$450 \gev \lsim \MA \lsim 1950 \gev$ and $6 \lsim \tb \lsim 10$.
However, if the higher $\Stopz$ mass is realized in nature,
corresponding to larger mixing in the $\Stop$ sector, in spite of the
relatively rough measurement of $\mstz$, in our example the allowed
parameter range is reduced to
$250 \gev \lsim \MA \lsim 1200 \gev$ and $2.5 \lsim \tb \lsim 3.5$.
Again, $\sweff$ plays an important role (cf.\ discussion of
\reffi{fig:MSt2MA}); without the
high precision measurement of $\sweff$ no upper limit on $\MA$ could
be set.

\smallskip
Thus the high precision measurements of $\MW$, $\sweff$ and $\Mh$ 
do not improve on the direct lower
bound on the mass of the pseudoscalar Higgs boson $A$. Instead they
enable us to set an {\em upper bound} on this basic parameter of the
supersymmetric Higgs sector, derived from the requirement of
consistency of the electroweak precision data with the MSSM.


\section{Conclusions}

The opportunity to measure electroweak observables very precisely
in the GigaZ mode of the prospective \epem\ linear collider TESLA, in
particular the electroweak mixing angle $\sweff$ and
the $W$~boson mass, opens new areas for high precision tests of electroweak
theories. We have analyzed in detail two generic examples:
(i) The Higgs mass of the Standard Model can be extracted to a precision
of a few percent from loop corrections. By comparison with the direct
measurements of the Higgs mass, bounds on new physics scales can be
inferred that may not be accessible directly.
(ii) The masses of particles in supersymmetric theories, which
for various reasons may not be accessible directly neither at the LHC nor at
the LC, can be constrained. Typical examples are the heavy Higgs bosons and the
heavy scalar top quark.
In the scenarios studied here, a sensitivity of up to order 2 TeV for the
mass of the pseudoscalar Higgs boson and an upper bound of 1 TeV for the
heavy scalar top quark could be 
expected from the virtual loop analyses of the high precision data.

Opening windows to unexplored energy scales renders these analyses of
virtual effects an important tool for experiments in the GigaZ mode of
a future \epem\ linear collider.


\section*{Acknowledgements}

We thank P.~Langacker for a critical reading of the manuscript and
A.~Hoang and T.~Teubner for helpful discussions.
Parts of the calculation have been performed on the QCM cluster at the
University of Karlsruhe, supported by the Deutsche
Forschungsgemeinschaft (Forschergruppe ``Quantenfeldtheorie und
Computeralgebra'').



\end{document}